\documentstyle[12pt]{article}

\input amssym.def
\input amssym
\font\teneufb=eufb10

\newfam\eufbfam
\textfont\eufbfam=\teneufb
\def\bfr#1{{\fam\eufbfam\relax#1}}

\input amssym

\title{ Smooth Loops and Thomas Precession}

\author{Alexander I. Nesterov \thanks{Departamento de F\'\i sica,
Universidad de Guadalajara, Guadalajara, Jalisco, M\'exico.  E-mail:
nesterov@udgserv.cencar.udg.mx. On leave from
Krasnoyarsk State University, Krasnoyarsk, Russia.} and Lev V. Sabinin
\thanks{Mathematics Institute, UNAM Morelia Branch, Mexico.  On leave
from Friendship of Nations University, Moscow, Russia}}

\begin{document}
\date{~}

\maketitle

%\newpage
\begin{abstract}
Fundamentals of the local smooth loops due to Sabinin are concisely
outlined together with the corresponding infinitesimal objects, so-called
$\mbox{\boldmath$\nu$}${\bf-hyperalgebras}, and the analogue of the Lie
groups theory. We apply here this theory to to formulation of a new concept
of loop of boosts. A quaternionic model of the three-parametric loop
of boosts is obtained and a remarkable connection with geodesic loops of
Lobachevskii space is found. A description of Thomas precession in the
light of general theory of smooth loops is given. \\

\noindent
Key words: Quasigroups, Loops, Thomas precession \hfill \\

\noindent
{MSC number(s): 20N05, 22E99, 51M15, 83A05  }
\end{abstract}

\newpage

\section{Introduction}

     In this article we discuss the problem of special Lorentz
transformations (hyperbolic rotations or boosts) and Thomas precession in
the light of general theory of smooth loop, see Sabinin (1972a,b, 1977,
1981, 1985, 1988, 1990, 1991, 1994). This theory is a direct generalization
of the Lie groups theory and it proved to be effective in applications
to Mathematical Physics. There is a lot of areas where this
theory has good promises, among of them Special and General Relativity,
Quantum Theory and Renormgroup Theory, see Nesterov (1989,1990), Kuusk et
al (1994).

The paper is 0rganized ed as following.

In Sec. 2 fundemantals of the local smooth loops theory are given,
the corresponding infinitisemal objects (so-called
$\mbox{\boldmath$\nu$}${\bf-hyperalgebras}) introduced and an analogue
of Lie groups theory is outlined.

In Sec. 3 the three-parametric loop of boosts is introduced, the
theory of smooth loops is used for its description and remarkable
connections with geodesic loops of  Lobachevskii space are traced.

In Sec. 4 we are treating the Thomas precession from the point of view of
smooth loops theory. Relevant loop $QH(2)$ and its properties are
considered.

In Sec. 5 concluding remarks are given.

\section{Smooth loops}

General results on the subject can be extract from Belousov (1967), Sabinin
(1972a,b, 1977, 1981, 1985, 1988, 1990, 1991, 1994). A {\bf quasigroup}
is a groupoid $\langle Q,\divideontimes\rangle$ in which
equations $a\divideontimes x=b,~y\divideontimes a=b$ have the unique
solutions:  $x=a\setminus b$, $y=b/a$. A {\bf loop} is a quasigroup with a
two-sided identity $a\divideontimes e= e\divideontimes a=a, \forall a \in
Q$. A loop $\langle Q,\divideontimes,e\rangle$ with a smooth functions
$\varphi(a,b):=a\divideontimes b$ is called the {\bf smooth loop}. Let
$\langle Q,\divideontimes,e\rangle$ be a smooth local loop with a neutral
element $e$. $Q$ being a manifold, dim $Q=n.$ We define
\begin{equation}
L_ab=a\divideontimes b=R_ba,\quad l_{(a,b)}=L^{-1}_{a\divideontimes
b}\circ L_a\circ L_b,
\end{equation}
where $L_a$ is a {\bf left translation}, $R_b$ is a {\bf right translation}
and $l_{(a,b)}$ is an {\bf associator}.

It is known that the infinitesimal theory of Lie groups arises from the
associativity of operation $\divideontimes$
\[
a\divideontimes(b\divideontimes c)=(a\divideontimes b)\divideontimes c.
\]
For the smooth quasigroups we have the modified law of
associativity ({\bf quasiassociativity})
\begin{equation}
a\divideontimes(b\divideontimes c)=(a\divideontimes b)\divideontimes l_{(a,b)}c.
\end{equation}
We will show below that the identity (2) leads to the infinitesimal theory
of smooth loops (Sabinin 1988, 1991).

{\bf Definition 2.1:} {\it The vector fields $A_\alpha\quad
(\alpha=1,\dots,n)$ defined on $Q$ are called the {\bf left fundamental
vector fields} of a local loop $\langle Q,\divideontimes,e\rangle$ if
\begin{equation}
A_\alpha(x)=A^\beta_\alpha(x)\frac{\partial}{\partial x^\beta},
\end{equation}
where
\[
A^\beta_\alpha(x)=\left[(L_x)_{\ast e}\right]^\beta_\alpha= \left[\frac{\partial (x\divideontimes
y)^\beta}{\partial y^\alpha}\right]_{y=e},
\]
\noindent
$\{x^\alpha\}$ being coordinates on $Q$}.

Analogously on can introduce the {\bf right fundamental vector fields}.
Obviously, $A_1,\dots,A_n$ are linearly independent at any point.

{\bf Definition 2.2:} {\it The differential 1-forms $\omega^\alpha\quad
(\alpha=1,\dots,n)$ of the base dual to $A_\alpha\quad (\alpha=1,\dots,n)$ are
called  the {\bf left fundamental 1-forms} of a loop $Q$.}

{\bf Definition 2.3}: {We define a {\bf quasialgebra} $\bfr q$ on $Q$
as the vector space spanned by left fundamental vector fields of $Q$ with
the Lie commutator as operation.}

Evidently,
\begin{equation}
[A_\alpha,A_\beta](x)=C^\gamma_{\alpha\beta}(x)A_\gamma(x).
\end{equation}

{\bf Theorem 2.1:} (Sabinin (1988, 1991)).
{\it Let $\langle Q, \divideontimes,e \rangle$ be a local loop. Then
$\varphi^\alpha = (a\divideontimes x)^\alpha$ and $\tilde l_\tau^\kappa =
\left [(l_{(a,x))\ast,e}\right ]^\kappa_\tau=\tilde l_\tau^\kappa(a,x)$
(see (1)) are the solutions of the system of differential equations
\begin{eqnarray}
\frac{\partial \varphi^\alpha}{\partial x^\mu} =
A_\gamma^\alpha(\varphi)\tilde l_\sigma^\gamma B_\mu^\sigma(x),
\nonumber\\ A_{\nu}^\sigma(x)\frac{\partial \tilde
l_\mu^{\kappa}}{\partial x^\sigma} - A_\mu^\sigma(x)\frac{\partial
\tilde l_{\nu}^{\kappa}}{\partial x^\sigma} =
C_{\nu\mu}^\sigma(x)\tilde l_\sigma^{\kappa} -
C_{\tau\lambda}^{\kappa}(\varphi)\tilde l_{\nu}^{\tau}\tilde l_\mu^\lambda
\end{eqnarray}
with initial conditions $\left.\varphi^\alpha\right|_e = a^\alpha$,
$\left.\tilde l_{\tau}^{\kappa}\right|_{e} = \delta_{\tau}^{\kappa}$
($\varphi^\alpha(a,e) = a^\alpha$,
$\tilde l_{\tau}^{\gamma}(a,e) = \delta_{\tau}^{\kappa}$).
Functions $A_{\gamma}^\alpha(x)$ are supposed to be given and satisfy the
conditions $A_{\gamma}^\alpha(x)(e) = \delta_{\gamma}^\alpha$;
$B_\mu^\sigma(x)$ is the inverse matrix for $A_{\beta}^\alpha(x)$ and
$C_{\nu\mu}^\sigma(x)$ are expressed through $A_{\beta}^\alpha(x)$ by
formula} (4).

{\it Remark 2.1}: One might exclude $\tilde l_{\nu}^{\kappa}$ from (5) and
obtain the system of second order in unknowns $\varphi^\alpha$.

 {\it Remark 2.2}: A solution of (5) is nonunique, there are some
arbitrariness, which will be seen further.

Let us introduce {\bf Exp} :$ T_{e}Q \rightarrow Q$ by
$\xi \rightarrow\varphi(1,\xi)$, where $\varphi(t,\xi)$ is the solution of the differential
equation
\begin{equation}
\frac{d\varphi^\alpha(t,\xi)}{dt} =
A_{\beta}^\alpha(\varphi(t,\xi))\xi^{\beta},\quad \varphi^\alpha(0,\xi)
= e^\alpha,\quad A_{\beta}^\alpha(e)=\delta_{\beta}^\alpha
\end{equation}
Here $A_{\beta}^\alpha(x)$ is supposed to be given. Due to the theorem
 of existance and uniqueness of solution it is easily verified that
$\varphi(t,\xi)=\varphi(1,t\xi) \stackrel{def}=\mbox{\bf Exp} (t\xi)$. It
is easy to see that $(\mbox{\bf Exp} )_{*,0} =$Id, what means that
$\mbox{\bf Exp}^{-1}$ locally exists.

 {\it Remark 2.3}: Taking into account that our construction
depends on the fields $\left[ A_\alpha(a)\right]^{\beta}
= A_\alpha^{\beta}(a)$,
we should write $\mbox{\bf Exp}(t\xi,A_1,\dots ,A_n)$ instead of $\mbox{\bf Exp} (t\xi)$,
but we shall do it in the case of misunderstanding only.

Now let us define in a neighbourhood of $e$ operations of a vector space,
Sabinin (1988, 1991)
\begin{equation}
x+y=\mbox{\bf Exp}(\mbox{\bf Exp}^{-1}x+\mbox{\bf
Exp}^{-1}y),\quad tx=\mbox{\bf Exp}(t\mbox{\bf Exp}^{-1}x) \quad (t\in \Bbb
R; x,y\in Q).  \end{equation} It is obvious that
\begin{equation}
\frac{d(tb)^\alpha}{dt}=A_{\beta}^\alpha(tb)(\mbox{\bf Exp}^{-1}b)^{\beta}.
\end{equation}
The operations $\omega_t : x\rightarrow tx$ can be added to an initial loop
$\langle Q,\divideontimes,e \rangle$ and we obtain the {\bf structure of a
canonical preodule of a given smooth loop}. All operations defined in (7)
can be added to an initial loop and we get the {\bf canonical prediodule}
of a given smooth loop.

{\it Remark 2.4:} The structure of any smooth odule (diodule), Sabinin
(1981, 1990, 1991), coincides with the canonical structure of its preodule
(prediodule).  Indeed, differentiating the identity of monoassociativity of
an odule:  $tb\divideontimes ub = (t+u)b$, by $u$ at $u=0$, we see that
$tb$ satisfies (13) with $\xi = \mbox{\bf Exp}^{-1}b$. Analogously,
$t(x+y)=tx+ty$ implies that after differentiating by $t$ \begin{equation}
A_{\beta}^\alpha(t(x+y))(\mbox{\bf Exp}^{-1}(x+y))^{\beta} =
A_{\beta}^\alpha(tx)(\mbox{\bf Exp}^{-1}x)^{\beta} +
A_{\beta}^\alpha(ty)(\mbox{\bf Exp}^{-1}y)^{\beta}.
\end{equation}
And at $t=0$ we have $\mbox{\bf Exp}^{-1}(x+y)=\mbox{\bf Exp}^{-1}x+\mbox{\bf Exp}^{-1}y$. Thus,
$x+y=\mbox{\bf Exp}(\mbox{\bf Exp}^{-1}x+\mbox{\bf Exp}^{-1}y))$.

One may now introduce (locally) for a smooth loop so called
{\bf normal coordinates} $b\rightarrow ((\mbox{\bf Exp}^{-1}b)^1, \dots,(\mbox{\bf Exp}^{-1}b)^n)$.
In such a coordinates $b^\alpha  = (\mbox{\bf Exp}^{-1}b)^\alpha$
$(\alpha =1,\dots ,n)$.

{\bf Definition 2.4:} (Sabinin (1988, 1991). {\it Let $V$ be a vector space
over $\Bbb R$ ($\dim V=n$), $d(\xi,\eta)$ be a continious binary operation
on $V$ , $d(\xi,\xi)=0$, admitting the representation
$d(\xi,\eta)=[d_{\alpha\beta}^{\gamma}(\xi)
\xi^\alpha\eta^{\beta}]e_{\gamma}$ in an arbitrary base $e_1,\dots,e_n$.
We say in the case, that $V$ is an {\bf hyperalgebra}. If on $V$,
additionally, a continuous binary operation $\nu(\eta,\xi)$, admitting
representation $\nu(\eta,\xi) = \nu_{\beta}^\alpha(\eta,\xi)\xi^\alpha$ in
an arbitrary base $e_1,\dots,e_n$ with the properties $\nu(0,\xi)=\xi$,
$\nu_{\beta}^\alpha(\eta,0)\xi^{\beta}=\xi^\alpha$ is given, then we say
 that $V$ is a $\nu$--{\bf hyperalgebra} (a hyperalgebra with a
multioperator $\nu$)}.

Evidently that {\bf any smooth loop $Q$ with the
neutral $e$ generates a $\mbox{\boldmath$\nu$}$-hyperalgebra on $T_{e}(Q)$ with
the operations} $d(\xi,\eta)=C_{\alpha\beta}^{\gamma}(\mbox{\bf Exp} \xi)\xi^\alpha\eta^{\beta}e_{\gamma}$,
$\nu(\eta,\xi)=\tilde l_{\beta}^\alpha(\mbox{\bf Exp} \eta,\mbox{\bf Exp} \xi)\xi^{\beta}e_\alpha$
(where $e_\alpha = (\partial\alpha)_{e}$). Such a $\nu$--hyperalgebra
is called {\bf the tangent $\mbox{\boldmath$\nu$}$--hyperalgebra} of a loop
$\langle Q,\divideontimes,e\rangle$ (see Sabinin (1988, 1991)).

Let us consider the system
\begin{equation}
\left. \begin{array}{l}
{\displaystyle \frac{d\varphi^\alpha}{dt}}=A_{\beta}^\alpha(\varphi)
 \nu^\beta(\mbox{\bf Exp}^{-1}a,t\mbox{\bf Exp}^{-1}b)t^{-1},  \\
\mbox{\hspace{-1.1in} where}\\
 A_{\beta}^\alpha(e)=\delta_{\beta}^\alpha,
 \quad \nu^{\beta}(\zeta,\eta) =
 \nu_{\gamma}^{\beta}(\zeta,\eta)\eta^{\gamma}, \\
\nu^{\beta}(0,\eta) =\eta,\quad
\nu^\beta_\gamma(\zeta,0)=\delta^\beta_\gamma, \quad \xi,\eta\in T_e(Q)
\end{array}\right\}
\end{equation}
with the initial conditions $\left. \varphi^\alpha\right|_{t=0}=a^\alpha$.
Here $\mbox{\bf Exp} \eta = \mbox{\bf Exp} (\eta,A_1,\dots ,A_n)$, $tb = \mbox{\bf Exp} (t\mbox{\bf Exp}^{-1}b)$ are defined
by given $A_{\beta}^i(x)$. The functions $ \nu^{\beta}$ are also
given. Note, that representation $\nu^{\beta}(\xi,\eta)$ through
$\nu_\alpha^{\beta}(\xi,\eta)$ is nonunique (it is possible that
$\nu^{\beta}(\xi,\eta) = \nu_{\gamma}^{\beta}(\xi,\eta)\eta^\gamma
 = \bar \nu_{\gamma}^{\beta}(\xi,\eta)\eta^\gamma).$

{ \bf Theorem 2.2:} (Sabinin (1988, 1991)). {\it Let a smooth local loop
$\langle Q,\divideontimes ,e \rangle$ together with its canonical
operations be given. Then $\varphi(a,b,t)=a\divideontimes tb$ is a solution
of the equation {\rm (10)}, where}
\begin{equation}
A_{\beta}^\alpha(a) = \left [
\left (L_a \right )_{*,e} \right ]_ {\beta}^\alpha,\quad
\nu^{\beta}(\xi,\eta) = \tilde l_{\gamma}^ {\beta}(\xi,\eta)\eta^\gamma
\end{equation}

{\it Proof:} Differentiating $(a\divideontimes tb)^\alpha$ by $t$ as
the composition of functions
and using (5) from  Theorem 2.1 and (10) we get our assertion.

{ \bf Theorem 2.3:} (Sabinin (1988, 1991)). {\it A solution
$\varphi^i(a,b,t)$ of the equation {\rm (10)} defines a local loop
$a\divideontimes b =\varphi(a,b,1)$ with the neutrul $e$ and its canonical
unary operations $tb=\varphi(e,b,t)$.  Moreover, $\varphi(a,b,t) =
a\divideontimes tb$.}

Let us introduce, Sabinin (1988, 1991)
\begin{equation}
\tilde A_{\alpha}^{\sigma}(x) = A_{\alpha}^{\lambda}(x)\frac{\partial(\mbox{\bf Exp}^{-1}x)^{\sigma}}{\partial x^{\lambda}}.
\end{equation}
Using (4)  we have
\begin{equation}
\tilde A_{\alpha}^{\sigma}(x)\frac{\partial\tilde A_{\beta}^{\gamma}(x)}{\partial (\mbox{\bf Exp}^{-1}x)^{\sigma}} -
\tilde A_{\beta}^{\sigma}(x)\frac{\partial\tilde A_{\alpha}^{\gamma}(x)}{\partial (\mbox{\bf Exp}^{-1}b)^{\sigma}} =
C_{\alpha\beta}^{\sigma}(x)\tilde A_{\sigma}^{\gamma}(x).
\end{equation}
Setting here $x=tb$ and contracting it with $(\mbox{\bf Exp}^{-1}b)^{\alpha}$ we get,
taking into account (6),
\begin{equation}
\frac{d\tilde A_{\beta}^{\gamma}(tb)}{dt} - \tilde A_{\beta}^{\sigma}(tb)t^{-1}[
\delta_{\delta}^{\gamma} - \tilde A_{\sigma}^{\gamma}(tb)]=
[C_{\alpha\beta}^{\sigma}(tb)(\mbox{\bf Exp}^{-1}b)^{\alpha}]\tilde A_{\sigma}^{\gamma}(tb)  .
\end{equation}
At the right hand side this equation has a singularity at $t=0$ ,
which does not allow us to use the theorem of existence and uniqueness.
Therefore, equivalently, we write an equation with matrix inverse to
$\tilde A_{\beta}^{\gamma}(x)$ :
\begin{equation}
\tilde B_{\mu}^{\lambda}(x) = B_{\nu}^{\lambda}(x)\frac{\partial x^{\nu}}
{\partial (\mbox{\bf Exp}^{-1}x)^{\mu}} .
\end{equation}
After evident calculations such an equation has the form
\begin{equation}
\frac{d[t\tilde B_{\mu}^{\lambda}(tb)]}{dt} = \delta_{\mu}^{\lambda}-
[C_{\alpha\mu}^{\sigma}(tb)(\mbox{\bf Exp}^{-1}b)^{\alpha}][t\tilde B_{\sigma}^{\lambda}(tb)].
\end{equation}
Thus, having solved the equation
\begin{equation}
\frac{d\psi_{\mu}^{\lambda}}{dt} = \delta_{\mu}^{\lambda}-
C_{\alpha\mu}^{\sigma}(tb)(\mbox{\bf Exp}^{-1}b)^{\alpha}\psi_{\sigma}^{\lambda},\quad
\left.\psi_{\sigma}^{\lambda}\right|_{t=0} = 0,
\end{equation}
we get
\begin{equation}
\tilde B_{\mu}^{\alpha}(b)= \psi_{\mu}^{\alpha}(1,b).
\end{equation}
Note, that $\lambda\psi_{\sigma}^{\alpha}(t,\lambda b) = \psi_{\sigma}^{\alpha}(\lambda t,b)$, being
solutions of the same equation coinciding at $t=0$. Thus, at $t=0$ we
receive $\psi_{\sigma}^{\alpha}(\lambda,b) =\lambda\psi_{\sigma}^{\alpha}(1,\lambda b)$ and the
solution takes the required form.

In such a way we can reconstruct $\tilde B_{\mu}^{\alpha}(x)$ and,
consequently, $\tilde A_{\mu}^{\alpha}(x)$ and  $A_{\mu}^{\alpha}(x)$ by
means of
\begin{equation}
d_{\mu}^{\sigma}(\xi) = C_{\alpha\mu}^{\sigma}(\mbox{\bf Exp}\xi)\xi^\alpha.
\end{equation}
In virtue of $C_{\alpha\mu}^{\sigma}=-C_{\mu\alpha}^{\sigma}$ it is evident,
that
\begin{equation}
d_{\mu}^{\sigma}(\xi)\xi^\mu=0.
\end{equation}
As for the rest, $d_{\mu}^{\sigma}(\xi)$ is an arbitrary function of the
form
\begin{equation}
d_{\mu}^{\sigma}(\xi)=d_{\alpha\mu}^{\sigma}(\xi)\xi^\alpha.
\end{equation}
(The representation of
$d_{\mu}^{\sigma}(\xi)$ in such a view is not unique, generally speaking.)
Having given an arbitrary $d_{\mu}^{\sigma}$ satisfying (20) and (21),
let us take invertible $\mbox{\bf Exp}$ arbitrarily . Let us introduce
$\bar tb=\mbox{\bf Exp}(t\mbox{\bf Exp}^{-1}b)$
and solve the equation
\begin{equation}
\frac{d\psi_{\mu}^{\lambda}}{dt} = \delta_{\mu}^{\lambda}- t^{-1}
d_{\mu}^{\sigma}(\mbox{\bf Exp}^{-1}\bar t b)\psi_{\sigma}^{\lambda},\quad
\left.\psi_{\sigma}^{\lambda}\right|_{t=0} = 0.
\end{equation}
We get $\tilde B_{\beta}^{\alpha}(\bar tb)=t^{-1}\psi_{\beta}^{\alpha}(t,b)$ and
$\tilde A_{\beta}^{\alpha}(\bar tb)$ (as an inverse matrix). Further,
using (13)--(17) we get
\begin{equation}
C_{\alpha\mu}^{\sigma}(b)({\mbox{\bf Exp}}^{-1}b)^{\alpha}=
d_{\mu}^{\sigma}(\mbox{\bf Exp}^{-1}b).
\end{equation}
And $d_{\mu}^{\sigma}$ has a prescribed meaning, that is by means of
arbitrary $d_{\mu}^{\sigma}$ we have reconstructed in the unique manner
$\tilde A_{\beta}^{\alpha}$ and , consequently , $ A_{\beta}^{\alpha}(x)$ for which
(23) is valid. The property (21) implies $\bar tb=tb$.  Indeed,
contracting (22) with $(\mbox{\bf Exp}^{-1}b)^{\mu}$ we get $d[t\tilde
 B_{\mu}^{\lambda}(\bar tb)(\mbox{\bf Exp}^{-1}b)^{\mu}]/dt = (\mbox{\bf
Exp}^{-1}b)^{\lambda}$ whence $\tilde B_{\mu}^{\lambda}(\bar tb)(\mbox{\bf
Exp}^{-1}b)^{\mu} = (\mbox{\bf Exp}^{-1}b)^{\lambda}$, or
\[
B_{\nu}^{\lambda}(\bar tb)\frac{\partial(\bar tb)^{\nu}}{\partial
(\mbox{\bf Exp}^{-1}\bar tb)^{\mu}}(\mbox{\bf Exp}^{-1}b)^{\mu}  =
(\mbox{\bf Exp}^{-1}b)^{\lambda},
\]
or
\begin{equation}
B_{\nu}^{\lambda}(\bar tb)\frac{\partial(\bar tb)^{\nu}}{\partial
\mbox{\bf Exp}^{-1}(\bar tb)^{\mu}}
\frac{d( \mbox{\bf Exp}^{-1}\bar tb)^{\mu}}{dt}=(\mbox{\bf
Exp}^{-1}b)^{\lambda} \end{equation} (we have used $\bar t = \mbox{\bf Exp}
t\mbox{\bf Exp}^{-1}$, or $\mbox{\bf Exp}^{-1}\bar tb= t\mbox{\bf
Exp}^{-1}b$).  Finally, $d(\bar tb)^{\nu}/dt = A_{\lambda}^{\nu}(\bar
tb)(\mbox{\bf Exp} ^{-1}b)^{\lambda}\quad \left. t b\right|_{t=0} =e)$,
which means $\bar tb =tb$. Since
 $d^\sigma(\xi,\eta)=d^\sigma_\mu(\xi)\eta^\mu$ we have

{\bf Theorem 2.4:} (Sabinin (1988, 1991)). {\it Any $\nu$--hyperalgebra $V$
with an operation $d$ uniquely defines a smooth loop $\langle
Q,\divideontimes,e\rangle$ with the tangent $\nu$--hyperalgebra isomorphic
to initially given.}

It is easy, also, to prove the following.

{\bf Theorem 2.5:} (Sabinin (1988, 1991)). {\it Any morphism of  smooth
loops induces a morphism of corresponding $\nu$--hyperalgebras and vice
versa.}

{\it Remark 2.5:} If operations  $\nu,d$ are analytic, then expanding
them into series one can introduce a countable system of multilinear
 operations (with identities), which is equivalent to initial $\nu$--
hyperalgebra (of course, some conditions of convergence are needed).

{\bf Theorem 2.6:} (Sabinin (1988, 1991)). {\it A smooth loop is
rightmonoalternative (that is $(x\divideontimes ty)\divideontimes
uy=x\divideontimes(t+u)y$)if and only if for its tangent
$\nu$--hyperalgebra $\nu(\eta,\xi)=\xi$ is valid (equivalently,
$\nu(\eta,\xi)$ is linear in $\xi$)}.

{\bf Theorem 2.7:} (Sabinin (1988, 1991). {\it If the tangent
$\nu$--hyperalgebra of a smooth loop is a Lie algebra, (that is an
operation $\xi\divideontimes \eta = d(\xi,\eta)$ are bilinear,
$\xi\divideontimes\xi =0,
\xi\divideontimes(\eta\divideontimes\zeta)+\zeta\divideontimes(\xi\divideontimes\eta)+
\eta\divideontimes(\zeta\divideontimes\xi)= 0$)\\
and this loop is rightmonoalternative (equivalently , $\nu(\eta,\xi)=\xi$)
then our loop is a Lie group. Converse is also true, since any Lie group is
rightmonoalternative.}

Note, that by definition a loop is {\bf monoalternative} if it is right and
left monoaltenative, that is $(x\divideontimes ty)\divideontimes uy=
x\divideontimes(t+u)y,\quad tx\divideontimes(ux\divideontimes )y= (t+u)x$.

In particular this is the case of smooth Moufang loops, Belousov (1967),
Sabinin (1985, 1990).

The following result is valid.

{\bf Theorem 2.8:} (Sabinin (1988, 1991)).{\it A local smooth
monoalternative loop is defined in the unique manner by its tangent
bilinear algebra.}

\section{Quaternions and smooth loops}

A quaternionic algebra over a field $\Bbb F$ is a set
\[
{\sf H} =
\{ \alpha + \beta i+ \gamma j+ \delta k \; \mid \;\alpha , \beta , \gamma ,
\delta \in {\Bbb F} \}
\]
with multiplication operation defined by the property of bilinearity and
following rules for $i, \; j, \; k$
\begin{equation}
i^{2}=j^{2}=k^{2} =
-1,\;\; jk=-kj = i,\;\;ki=-ik=j,\;\; ij=-ji= k.
\end{equation}

We consider a quaternionic algebra over complex field ${\Bbb C}(1,\rm i)$
\[
{\sf H}_{\cal C} = \{ \alpha + \beta i+ \gamma j + \delta k\; \mid
 \; \alpha , \beta,\gamma,\delta \in {\Bbb C} \}.
\]
The quaternionic conjugation (denoted by $^{+}$) is defined by
\begin{equation}
q^+=\alpha - \beta i -\gamma j - \delta k.
\end{equation}
for $q=\alpha + \beta i+ \gamma j + \delta k$.
This definition implies
\begin{equation}
(q p)^+ = p^+ q^+, \quad p,q\in{\sf H_{\cal C}}.
\end{equation}
For real quaternions $q\in \sf H_{\cal R}$:
\begin{equation}
q=\alpha + \beta i+ \gamma j+ \delta k;\;
\alpha , \beta , \gamma , \delta \in {\Bbb R},
\end{equation}
we define the norm $\|q\|^2$ setting
\begin{equation}
\|q_ r\|^2=qq^+=\alpha^2 + \beta^2+ \gamma^2+ \delta^2.
\end{equation}

Let us consider a set of unit quaternions ${\sf H}_{\cal I}$:
\begin{equation}
{\sf H}_{\cal I} =\{q=\alpha + {\rm i}(\beta i+ \gamma j+ \delta
k):\; \|q\|^2=1,\; \;{\rm i}^2=-1,\;\rm i\in{\Bbb C};\; \alpha ,
\beta , \gamma , \delta \in {\Bbb R} \},
\end{equation}
where the norm $\|q\|^2$ is given by
\begin{equation}
\|q\|^2=q {q}^+=\alpha^2 - \beta^2- \gamma^2-
\delta^2.
\end{equation}

It is easy to see that for $p,\, q \in{\sf
H}_{\cal I}$ the product $p q \not\in{\sf H}_{\cal I}$.
It means that the set ${\sf H}_{\cal I}$ is not closed with respect to
the multiplication of quaternions. We introduce a new
operation $\circledast$ such one that $p\circledast q
\in{\sf H}_{\cal I}$. This new operation is defined by
\begin{equation}
p\circledast q = p q
{l}^+,
\end{equation}
where $p, q \in {\sf H}_{\cal I},\quad l \in {\sf H}_{\cal
R}:  \|p\|^2=\|q\|^2=\|l\|^2=1$.  It leads to
$\|p\circledast q\|^2=1$ as well. Requiring
$p\circledast q \in{\sf H}_{\cal I}$, we shall find
a quaternion $l$.

In order to simplify following considerations let us set
\begin{equation}
p=\zeta_0(1+{\rm i}\zeta), \quad
q =\eta_0(1+{\rm i}\eta), \quad  l= \alpha_0(1+\alpha),
\end{equation}
where
\[
\zeta=\zeta^1 i+\zeta^2 j+\zeta^3 k, \quad
\eta=\eta^1 i+\eta^2 j+\eta^3 k, \quad
\alpha=\alpha^1 i+\alpha^2 j+\alpha^3 k, \quad
(\alpha^i,\;\eta^j,\;\zeta^k\in \Bbb R),
\]
\[
\zeta_0=\frac{1}{\sqrt{1-\mbox{\boldmath{$\zeta\zeta$}}}},\quad
\eta_0=\frac{1}{\sqrt{1-\mbox{\boldmath{$\eta\eta$}}}},\quad
\alpha_0=\frac{1}{\sqrt{1+\mbox{\boldmath{$\alpha\alpha$}}}},
\]
and ${\mbox{\boldmath$\zeta$}}=(\zeta^1,\;\zeta^2,\;\zeta^3),\quad
\mbox{\boldmath$\zeta\zeta$}=(\zeta^1)^2+(\zeta^2)^2+(\zeta^3)^2,$  etc.

Requiring  $p\circledast q \in{\sf H}_
{\cal I}$, we find from (32)
\begin{equation}
l=\frac{1+\zeta^+\eta} {\|1+\zeta^+\eta\|}.
\end{equation}
Finallly, using (32) and (34) we obtain
\begin{equation}
p\circledast q=\frac{\|1+\zeta^+\eta\|}
{\sqrt{(1-\mbox{\boldmath{$\zeta\zeta$}})(1-\mbox{\boldmath{$\eta\eta$}})}}
\left(1+{\rm
i}\frac{(\zeta+\eta)(1+\eta^+\zeta)}{\|1+\zeta^+\eta\|^2}\right),
\end{equation}
and obviously $p\circledast q\in {\sf H}_{\cal I}$.

It easy to see that the set of unit quaternions ${\sf H}_{\cal I}$ with the
nonassociative operation (35) is a loop. We call this loop
$Q{\sf H}_{\cal I}$. The loop $Q{\sf H}_{\cal I}$ forms so-called
non-associative quaternionic representation and  evidently it is isomorphic
to the loop $QH(3)$, which we define in open ball
$D_3=\{\mbox{\boldmath{$\zeta$}}:  |\mbox{\boldmath{$\zeta$}}|<1\}$ by
\begin{equation}
\zeta\divideontimes\eta=(\zeta+\eta)/(1+\zeta^+\eta),
\end{equation}
where $\zeta=\alpha i+\beta j +\gamma k \leftrightarrow
\mbox{\boldmath{$\zeta$}}=(\alpha,\beta,\gamma)$ and
$/$ denotes the right division. The associator (see (1)) is given by
\begin{equation}
l_{(\zeta,\eta)}\xi= (1+\eta^+\zeta)\setminus(1+\zeta\eta^+)\xi,
\end{equation}
where $\setminus$ is the left division. This expression one can write as
\begin{equation}
l_{(\zeta,\eta)} \xi= \frac{1+\mbox{\boldmath{$\zeta\eta$}}-|\mbox
{\boldmath{$\zeta$}}\times\mbox{\boldmath{$\eta$}}|^2}
{1+\mbox{\boldmath{$\zeta\eta$}}
+|{\mbox{\boldmath{$\zeta$}}}\times{\mbox{\boldmath{$\eta$}}}|^2} \xi.
\end{equation}

In the vector form the operation (36) is written as
\begin{equation}
\mbox{\boldmath$\zeta$}\divideontimes\mbox{\boldmath$\eta$}=
\frac{1+2\mbox{\boldmath$\zeta$}\mbox{\boldmath$\eta$}
+\mbox{\boldmath$\eta$}^2}{1+2\mbox{\boldmath$\zeta$}\mbox{\boldmath$\eta$}
+\mbox{\boldmath$\zeta$}^2\mbox{\boldmath$\eta$}^2}\mbox{\boldmath$\zeta$}
+\frac{1-\mbox{\boldmath$\zeta$}^2}
{1+2\mbox{\boldmath$\zeta$}\mbox{\boldmath$\eta$}
+\mbox{\boldmath$\zeta$}^2\mbox{\boldmath$\eta$}^2}\mbox{\boldmath$\eta$}
\end{equation}

{\it Remark 3.1:} Using the identity
\[
(\zeta+\eta)(1+\eta^+\zeta)=(1+\zeta^+\eta)(\zeta+\eta),
\]
where $\zeta=\alpha i+\beta j +\gamma k,\quad \eta=\lambda i +\mu j +
\nu k$, one can write (36) as
\begin{equation}
\zeta\divideontimes\eta=(1+\eta^+\zeta)\setminus(\zeta+\eta),
\end{equation}

{\it Remark 3.2:} The loop $QH(3)$ is isomorphic to geodesic loops of
three-dimensional Lobachevskii space realized as the upper part of
two-sheeted unit hyperboloid $H^3$. The isomorphism is established by
exponential mapping (see Eq. (47))
\[
\mbox{\boldmath$\zeta$}= {\mbox{\bf Exp}}
\mbox{\boldmath$\tau$}=\frac{\mbox{\boldmath$\tau$}}{\tau}
\tanh\tau, \quad
\mbox{\boldmath$\tau$}= {\mbox{\bf Exp}^{-1}}
\mbox{\boldmath$\zeta$}=\frac{\mbox{\boldmath$\zeta$}}
{|\mbox{\boldmath$\zeta$}|}
\tanh^{-1}|\mbox{\boldmath$\zeta$}|,
\]
where we introduced $\tau\equiv|\mbox{\boldmath$\tau$}|
= \tanh^{-1}|\mbox{\boldmath$\zeta$}|$.

Note also, that in this case it is valid so-called {\bf identity of
pseudolinearity}, Sabinin  et al (1986), Sabinin and Miheev (1993):
\begin{equation}
\mbox{\boldmath$\zeta$}\divideontimes\mbox{\boldmath$\eta$}=
\mbox{\bf Exp}(\alpha(\mbox{\boldmath$\zeta,\eta$})
\mbox{\bf Exp}^{-1}\mbox{\boldmath$\zeta$}+
\beta(\mbox{\boldmath$\zeta,\eta$})
\mbox{\bf Exp}^{-1}\mbox{\boldmath$\eta$}),
\end{equation}
where
\[
\alpha(\mbox{\boldmath$\zeta,\eta$})=
\left(\frac{1+2\mbox{\boldmath$\zeta$}\mbox{\boldmath$\eta$}
+\mbox{\boldmath$\eta$}^2}{1+2\mbox{\boldmath$\zeta$}\mbox{\boldmath$\eta$}
+\mbox{\boldmath$\zeta$}^2\mbox{\boldmath$\eta$}^2}\right)
\left(\frac{|\mbox{\boldmath$\zeta$}|\tanh^{-1}
{|\mbox{\boldmath$\zeta$}\divideontimes\mbox{\boldmath$\eta$}|}}
{{|\mbox{\boldmath$\zeta$}\divideontimes\mbox{\boldmath$\eta$}|}
\tanh^{-1}{|\mbox{\boldmath$\zeta$}|}}\right),
\]
\[
\beta(\mbox{\boldmath$\zeta,\eta$})=
\left(\frac{1-\mbox{\boldmath$\zeta$}^2}
{1+2\mbox{\boldmath$\zeta$}\mbox{\boldmath$\eta$}
+\mbox{\boldmath$\zeta$}^2\mbox{\boldmath$\eta$}^2}\right)
\left(\frac{|\mbox{\boldmath$\eta$}|\tanh^{-1}
{|\mbox{\boldmath$\zeta$}\divideontimes\mbox{\boldmath$\eta$}|}}
{{|\mbox{\boldmath$\zeta$}\divideontimes\mbox{\boldmath$\eta$}|}
\tanh^{-1}{|\mbox{\boldmath$\eta$}|}}\right),
\]

It is easy to verify (see Belousov (1967), Sabinin (1990, 1991), Sabinin and
Miheev (1993) that
\begin{equation}
\xi\divideontimes(\eta\divideontimes(\xi\divideontimes\zeta))=
(\xi\divideontimes(\eta\divideontimes\xi))\divideontimes\zeta \qquad
\mbox {(left Bol identity)},
\end{equation}
and
\begin{equation}
\xi\divideontimes(\eta\divideontimes(\eta\divideontimes\xi))=
(\xi\divideontimes\eta)\divideontimes(\xi\divideontimes\eta) \qquad
\mbox {(left Bruck identity)}.
\end{equation}

Let us consider the infinitesimal theory of the loop $QH(3)$. From (39) we
obtain the {\bf left fundamental vector fields} of the loop $QH(3)$:
\begin{equation}
A_i=(1-\mbox{\boldmath{$\zeta\zeta$}})\frac{\partial}{\partial
\zeta^i} .
\end{equation}
These vector fields obey the commutation relations of quasialgebra
\begin{equation}
[A_i,A_j]=C^k_{ij}(\zeta)A_k,
\end{equation}
where
\[
C^k_{ij}(\zeta)=2(\delta^k_i\zeta_j-\delta^k_j\zeta_i)
\]
are the structure functions and $\zeta^i=\zeta_i$. Dual base of 1-forms are
determined by
\begin{equation}
\omega^i=\frac{d\zeta^i}{1-\mbox{\boldmath{$\zeta\zeta$}}}.
\end{equation}

Now let us find the exponential mapping
$T_eQ \stackrel{\mbox{\bf Exp}}{\longrightarrow} Q$ by solving the Eq. (6).
Using (44), in normal coordinates $\tau^i=\tau n^i$
($|\mbox{\boldmath{$n$}}|=1$), one can write (6) as
\[
\frac{d\varphi^i}{d t}=(1-\Sigma_{j=1}^3\varphi^j\varphi^j)n^i.
\]
The solution of this system is $\varphi^i=n^i\tanh t$. Thus the exponential
mapping is given by
\begin{equation}
\mbox{\boldmath$\zeta$}= {\mbox{\bf Exp}}
\mbox{\boldmath$\tau$}=\frac{\mbox{\boldmath$\tau$}}{\tau}
\tanh\tau, \quad
\mbox{\boldmath$\tau$}= {\mbox{\bf Exp}^{-1}}
\mbox{\boldmath$\zeta$}=\frac{\mbox{\boldmath$\zeta$}}
{|\mbox{\boldmath$\zeta$}|}
\tanh^{-1}|\mbox{\boldmath$\zeta$}|,
\end{equation}

Let us consider the two-parametric subloop $QH(2)\subset QH(3)$ which we
define in the open disk
$D_2=\{\mbox{\boldmath{$\zeta$}}:  |\mbox{\boldmath{$\zeta$}}|<1\}$ by
\[
\zeta\divideontimes\eta=(\zeta+\eta)/(1+\zeta^+\eta),
\]
where $\zeta=\alpha i+\beta j \Leftrightarrow
\mbox{\boldmath{$\zeta$}}=(\alpha,\beta)$. The one of the remarkable
peculiaritis of this loop is that it can realized  on the complex
numbers.  Let $\Bbb C$ be the complex plane, and let $D_2\subset\Bbb C$ be
the open unit disk, $D_2=\{\zeta\in\Bbb C:\left|\zeta\right|<1 \}$. Inside
the disk $D_2$ we define the binary operation
\begin{equation}
L_\zeta\eta\equiv\zeta\divideontimes\eta=\frac{\zeta+\eta}{1+\zeta^\ast\eta},
\end{equation}
where $\zeta,\eta\in D$ and $\zeta^\ast$ is the complex conjugate number
$(\zeta=\zeta^1+\rm i \zeta^2, \quad \zeta^\ast=\zeta^1-\rm i \zeta^2)$.

The set of complex numbers with the operation $\divideontimes$ on $D_2$
forms two-sided loop $QH(2)$, Nesterov and Stepanenko (1986),
Nesterov (1989, 1990).  The associator
$l_{(\zeta,\eta)}=L^{-1}_{\zeta\divideontimes\eta}\circ L_\zeta\circ
L_\eta$ on $QH(2)$ is determined by
\begin{equation}
l_{(\zeta,\eta)}\xi=\frac{1+\eta\zeta^\ast}{1+\eta^\ast \zeta}\xi
\equiv e^{-\rm i\varphi}\xi, \quad \varphi=2\mbox{arg}(1+\eta^\ast \zeta).
\end{equation}

This loop is isomorphic to the geodesic loop of two-dimensional
Lobachevskii space realized as the upper part of two-sheeted unit
hyperboloid $H^2$.The isomorphism is established by exponential mapping
\begin{equation}
\zeta=e^{\rm i\varphi}\tanh\frac{\theta}{2} ,
\end{equation}
where $(\theta,\varphi)$ are inner coordinates  of unit $H^2$.

\section{Loop of boosts and Thomas precession}

The set of matrices
\begin{equation}
U_{\mbox{\boldmath$\theta$}}= \cosh{(\theta/2)}+
(\mbox{\boldmath$n$}\cdot \mbox{\boldmath$\sigma$})
\sinh{(\theta/2)},\quad |\mbox{\boldmath$n$}|=1.
\end{equation}
determines in the Minkowski space hyperbolic rotations (boosts) (see,
e.g., Misner et al (1973))  by
\begin{equation} X'=UXU^\ast,
\end{equation}
where the matrix $X$ is connected with four-vector $x^\mu$ by
\[
X=x^0+x^i\sigma_i.
\]
The angle $\theta$ is related with the velocity of
the  reference system by $\beta=\tanh \theta$, where $\beta=v/c$ and $c$ is
the speed of light. The unit vector $\mbox{\boldmath$n$}$ determines the 
direction of boost. It is well known that the set of hyperbolic rotations 
does not form the group.  It forms the loop  with the following 
nonassociative operation, Nesterov (1989, 1990) 
\begin{eqnarray} 
U_{\mbox{\boldmath$\theta$}_1} \circledast
U_{\mbox{\boldmath$\theta$}_2}=
U_{\mbox{\boldmath$\theta$}_1}U_{\mbox{\boldmath$\theta$}_2}
\Lambda^{-1}_{(\mbox{\boldmath$\theta$}_1,\mbox{\boldmath$\theta$}_2)},
\end{eqnarray}
where $\mbox{\boldmath$\theta$}=\theta\mbox{\boldmath$n$},\quad
\left|\mbox{\boldmath$n$}\right|=1$ and
\begin{eqnarray}
U_{\mbox{\boldmath$\theta$}_1} =\cosh(\theta_1/2)+
(\mbox{\boldmath$n$}_1\cdot\mbox{\boldmath$\sigma$})
\sinh(\theta_1/2),\\
U_{\mbox{\boldmath$\theta$}_2} =\cosh(\theta_2/2)+
(\mbox{\boldmath$n$}_2\cdot\mbox{\boldmath$\sigma$}) \sinh(\theta_2/2),\\
\Lambda_{(\mbox{\boldmath$\theta$}_1,\mbox{\boldmath$\theta$}
_2)}=\cos(\alpha/2)+
\rm i(\mbox{\boldmath$n$}_\alpha\cdot\mbox{\boldmath$\sigma$}) \sin
(\alpha/2),\\
\cot(\alpha/2)=\frac
{\mbox{\boldmath$n$}_1\cdot\mbox{\boldmath$n$}_2 +
\cot(\theta_1/2)\cot(\theta_2/2)}
{\sqrt{1-(\mbox{\boldmath$n$}_1\cdot\mbox{\boldmath$n$}_2)^2}},\\
\mbox{\boldmath$n$}_\alpha=\frac{\mbox{\boldmath$n$}_1\times
\mbox{\boldmath$n$}_2 }
{\sqrt{1-(\mbox{\boldmath$n$}_1\cdot\mbox{\boldmath$n$}_2)^2}},
%{\left|\mbox{\boldmath$n$}_1\times\mbox{\boldmath$n$} _2\right|}.
\end{eqnarray}

Now let us consider the quaternionic formulation of boosts. The space-time
points can be represented by quaternions as follows
\begin{equation}
X=t+{\rm i}(xi +yj+zk).
\end{equation}
and the Lorentz invariant norm is given by
\begin{equation}
XX^+=t^2-x^2-y^2-z^2.
\end{equation}

Then  the special Lorentz transformations (boosts), which are determined
by (53), can be represented by the action of the linear quaternions
\[
Q =\zeta_0(1-{\rm i}\zeta), \quad Q\in Q{\sf H}_{\cal I}
\]
on the quaternions
\[
X=t-{\rm i}(xi +yj+zk),
\]
namely,
\[ X'=Q X {Q}^+.
\]
The three-parametric loop of boosts is isomorhpic to the loop
$Q{\sf H}_{\cal I}$. This isomorphism is established by
\begin{eqnarray}
\mbox{\boldmath$\zeta=
n$}\tanh\frac{\theta}{2} ,\quad |\mbox{\boldmath$n$}|=1, \\
i\longrightarrow {\rm i}\sigma_z,\quad
j\longrightarrow {\rm i}\sigma_y,\quad
k\longrightarrow {\rm i}\sigma_x,\quad
\end{eqnarray}
where $\sigma_x,\;\sigma_y,\;\sigma_z,\;$  are Pauli matrices. Hence, we
have
\begin{equation}
Q\longrightarrow U= \cosh{(\theta/2)}+
(\mbox{\boldmath$n$}\cdot \mbox{\boldmath$\sigma$})
\sinh{(\theta/2)}.
\end{equation}

Setting $\beta=\zeta$ (or $\mbox{\boldmath$\beta$}=\mbox{\boldmath$\zeta=
n$}\tanh\frac{\theta}{2}$, where $\mbox{\boldmath$\beta$}$ is the
three-velocity of the observer, the speed of light $c=1$), we obtain from
(36) the quaternionic formula for the addition of relativistic
three-velocities
\begin{equation}
\beta_1\divideontimes\beta_2=(\beta_1+\beta_2)/(1+{\beta_1}^+\beta_2),
\end{equation}
or equivalently
\begin{equation}
\beta_1\divideontimes\beta_2=(1+{\beta_2}^+\beta_1)\setminus
(\beta_1+\beta_2).
\end{equation}

Now let us consider the loop $QH(2)$  with nonassociative operation
$\divideontimes$ (see(48))
\begin{equation}
L_\zeta\eta\equiv\zeta\divideontimes\eta=\frac{\zeta+\eta}{1+\zeta^\ast\eta}.
\end{equation}
We assign to each element $\zeta\in QH(2)$ the matrix
$U_\zeta\in SU(1,1)$:
\begin{equation}
\zeta\longrightarrow U_\zeta=
\left(\begin{array}{cc}
a & b\\
b^\ast & a
\end{array}\right),\quad a^2-\left|b\right|^2=1,
\end{equation}
where
\[
a=\frac{1}{\sqrt{1-\left|\zeta\right|^2}}, \quad
b=\frac{\zeta}{\sqrt{1-\left|\zeta\right|^2}}
\]
and define the nonassociative operation $\circledast$ on the set $\sf K$
of matrices (67) as
\begin{equation}
U_\eta\circledast U_\zeta=U_{\eta\divideontimes\zeta}.
\end{equation}
Note, that
\begin{equation}
U_{\eta\divideontimes \zeta}=U_\eta U_\zeta\Lambda^{-1}(\eta,\zeta),
\end{equation}
where
\begin{equation}
\Lambda=
\left(\begin{array}{cc}
e^{-i\varphi/2} & 0\\
0 & e^{i\varphi/2}
\end{array}\right),\quad \varphi=2\mbox{arg}(1+\eta^\ast\zeta).
\end{equation}

The set $\sf K$ with the composition law $\circledast$
forms the so-called the {\bf matrix representation} of the loop $QH(2)$.

Let $\eta=\nu\tanh{(\theta/2)}$, where $\left|\nu\right|=1$. Then
\[
a=\frac{1}{\sqrt{1-\left|\zeta\right|^2}}=\cosh{(\theta/2)}, \quad
b=\frac{\zeta}{\sqrt{1-\left|\zeta\right|^2}}=\nu\sinh{(\theta/2)},
\]
and hence
\[
\zeta\longrightarrow U_{\mbox{\boldmath$\theta$}}= \cosh{(\theta/2)}+
(\mbox{\boldmath$\nu$}\cdot \mbox{\boldmath$\sigma$})
\sinh{(\theta/2)}=\exp(\frac{1}{2}\mbox{\boldmath$\theta$}\cdot
\mbox{\boldmath$\sigma$}),
\]
where
$\mbox{\boldmath$\theta$}=\theta\mbox{\boldmath$\nu$},
\quad \mbox{\boldmath$\nu$}=(\Re\nu,\ \Im\nu)$ and $\sigma_i$ are Pauli
matrices .

Let us consider the combination of two boosts in different
directions but with the same angle $\theta$:
\begin{equation}
\eta=\nu \tanh \frac{\theta}{2},\quad \zeta=(\nu+\delta\nu)\tanh \frac{\theta}{2}.
\end{equation}
Setting $\nu=\exp(\rm i \alpha)$ one can
obtain for the infinitesimal $\delta\varphi$ (see (49), (70)) the following
expression:
\begin{equation}
\delta\varphi=2\frac{\delta\alpha\tanh^2\frac{\theta}{2}}
{1+\tanh^2\frac{\theta}{2}}.
\end{equation}

Let $\delta\varphi=\omega d t$, then (72) leads to
\begin{equation}
\omega=2\frac{\stackrel{\centerdot}\alpha\tanh^2\frac{\theta}{2}}
{1+\tanh^2\frac{\theta}{2}}.
\end{equation}
For the slow motion ($\theta\ll 1$) one can easy obtain
\begin{equation}
\omega=\frac{1}{2}\theta^2\stackrel{\centerdot}\alpha =\frac{1}{2}a v,
\end{equation}
where $a=v^2/r$ is the acceleration , $v$ being the
velocity of the reference system. In an arbitrary  coordinate system (73)
takes the following form
\begin{equation}
\mbox{\boldmath$\omega$}=\frac{1}{2}\mbox{\boldmath$a$} \times
\mbox{\boldmath$v$}.
\end{equation}
The last one is exactly the expression for Thomas precession (see,
e.g. Misner et al (1973)). We see that associator $l_{(\zeta.\eta)}$
completely determines the Thomas precession.

\section{Concluding remarks}

For the first time the complex model of relativistic addition of
velocities was introduced by Nesterov (Nesterov and Stepanenko (1986),
Nesterov (1989, 1990)) and later by Ungar (1991a,b, 1992,
1994)) . The other models of addition of three-velocities was 
considered by Sabinin and Miheev (1993). So-called ``gyrogroup'', see 
Ungar (1991a), is exactly left Bol loop with left Bruck identity, 
Sabinin (1995).  Some vectorlike properties of the complex disk, Ungar 
(1994), are contained in the vector space operations (7).  In particular, 
Eq.(8.4b) in Ungar (1991a) then expresses obvious properties of the 
exponential mapping (50), and some of axioms of gyrogroups are superfluous, 
Sabinin (1995).  The so-called gyrosemidirect product, Ungar (1991a), 
proved to be a subcase of the well known semidirect product of a loop by 
its transassociant, Sabinin (1972a,b).  The fact that several authors 
independly rediscovered certain results of the theory of smooth loops in 
connection with some physical problems, is remarkable and valuable, showing 
the vitality and importance of the smooth loops theory.

\section*{Acknowledgements}

We are greatly indebted to Professor B.N. Apanasov (Oklahoma University,
USA) and Professor L.L. Sabinina (Michoacan University, Mexico) who have
attracted our attention to publications on the subject. We would
like to express our gratitude to Professor L.V. Sbitneva (Moscow Institute
of Electronics and Mathematics, Russia) for valuable advices in the process
of writing this article.

\newpage
\section*{References}
\begin{description}

\item[] {} Belousov, V.D.: 1967
Foundation of quasigroups and loops theory. Moscow: 1967

\item[]{} Kuusk, P., \"Ord, J., Paal, E.:
J. Math. Phys. {\bf 31}, 321 (1994)

\item[]{} Misner, C.W., Thorne, K.S., Wheeler, J.A.:
{ Gravitation.} San Fransisco: W.H. Freeman 1973

\item[] {} Nesterov, A.I., Stepanenko, V.A.:
 On methods of nonassociative algebra in geometry and physics.
Preprint 400 F. Krasnoyarsk: L. V. Kirensky Institute of Physics 1986

\item[] {} Nesterov, A.I.:
{ Methods of nonassociative algebra in physics}
Dr.Sci. Thesis. Tartu: Institute of Physics 1989

\item[] {} Nesterov, A.I.:
{ Quasigroup ideas in physics}; in { Quasigroups and nonaassociatives
algebras in physics.} Tartu: Institute of Physics 1990

\item \label{Sabinin 1972}
Sabinin, L.V.:
{  On the equivalence of categories of loops and homogeneous spaces.}
Soviet Math. Dokl.
{\bf 13}, 970 (1972a)

\item[] {}Sabinin, L.V.: The geometry of loops.
 Mathematical Notes.
{\bf 12}, 799 (1972b)

\item[] {}Sabinin, L.V.
{  Odules as a new approach to a geometry with a connection.}
Soviet Math. Dokl.
{\bf 18}, 515 (1977)

\item[] {}Sabinin, L.V.:
{ Methods of Nonassociative Algebra in Differantial Geometry},
in: { Suppliment to Russian translation of S.K. Kobayashy and K. Nomizy
``Foundations of Differential Geometry'', Vol. 1.} Moscow: Nauka 1981

\item[] {}Sabinin, L.V.:
{  The Theory of Smooth Bol loops.}
Moscow: Friendship of Nations University Press 1985

\item[] {}Sabinin, L.V.:
{ Differential equations of smooth loops}, in:
 { Proceedings of Sem. on Vector and Tensor Analysis}
{\bf 23}, 133. Moscow: Moscow Univ. Press 1988

\item[] {}Sabinin, L.V.:
{ Quasigroups and Differential Geometry},
in { Quasigroups and Loops: Theory and Applications},
edited by O. Chein, H. Pflugfelder and J. D. H. Smith.
Berlin: Heldermann Verlag 1990

\item[]{} Sabinin, L.V.:
{  Analytic Quasigroups and Geometry.}
Moscow: Friendship of Nations University 1991

\item[] {}Sabinin, L.V.:
{ On differential equations of smooth loops.}
 Russian Mathematical Survey.
{\bf 49}, 172 (1994)

\item[] Sabinin, L.V., Miheev, P.O.:
{ On the law of addition of velocities in Special Relativity.}
 Russian Mathematical Survey,
{\bf 48}, N 183 (1993)

\item[] {}Sabinin, L.V., Matveev, O.A.,
Yantranova, S.S.: On the identity of quasi-linearity in
differentiable linear geodiodular manifolds, in: { Invariant tensors}
Moscow: VINITI, Dep No 6553-B86 (1986)

\item[] {} Sabinin, L.V.:
{ On gyrogroups of A. Ungar}.
 Russian Mathematical Survey,
{\bf 50 (2)} (1995)

\item[]{} Ungar, A.A.:
{ Thomas precession and its associated grouplike structure},
Am. J. Phys.
{\bf 59}, 824 (1991a)

\item[]{} Ungar, A.A.:
{ The abstract  complex Lorentz transformation group with real metric.
I. Special relativity formalism to deal with the holomorphic
automorphism group of the unit ball in any complex Hilbert space.}
J. Math. Phys.
{\bf 35}, 1408 (1991b)

\item[]{} Ungar, A.A.:
{ The abstract Lorentz transformation group.}
Am. J. Phys.
{\bf 60}, 815 (1992)

\item[]{} Ungar, A.A.
{ The holomorphic automorphism group of the complex disk.}
Aeq. Math.
{\bf 47}, 240 (1994)

\end{description}

\end{document}